\documentclass[superscriptaddress, prl]{revtex4}

\newcommand{\kB}{k_\mathrm{B}}

\usepackage{graphicx}

\begin{document}

\preprint{}
 
\title{Information heat engine: converting information to energy by feedback control}

\author{Shoichi Toyabe}
\affiliation{Department of Physics, Faculty of Science and Engineering, Chuo University, Kasuga, Tokyo 112-8551, Japan}
\author{Takahiro Sagawa}
\affiliation{Department of Physics, Graduate School of Science, University of Tokyo, Hongo, Tokyo 113-0033, Japan}
\author{Masahito Ueda}
\affiliation{Department of Physics, Graduate School of Science, University of Tokyo, Hongo, Tokyo 113-0033, Japan}
\affiliation{ERATO Macroscopic Quantum Control Project, JST, Yayoi, Tokyo 113-8656, Japan}
\author{Eiro Muneyuki}
\affiliation{Department of Physics, Faculty of Science and Engineering, Chuo University, Kasuga, Tokyo 112-8551, Japan}
\author{Masaki Sano}
\affiliation{Department of Physics, Graduate School of Science, University of Tokyo, Hongo, Tokyo 113-0033, Japan}

\begin{abstract}
In 1929, Leo Szilard invented a feedback protocol\cite{Szilard1929} in which a hypothetical intelligence called Maxwell's demon pumps heat from an isothermal environment and transduces it to work.
After an intense controversy that lasted over eighty years; it was finally clarified that the demon's role does not contradict the second law of thermodynamics, implying that we can convert information to free energy in principle\cite{Landauer1961, Bennet1982, Sagawa2009, MaxwellDemon2, Maruyama2009}.
Nevertheless, experimental demonstration of this information-to-energy conversion has been elusive.
Here, we demonstrate that a nonequilibrium feedback manipulation of a Brownian particle based on information about its location achieves a Szilard-type information-energy conversion.
Under real-time feedback control, the particle climbs up a spiral-stairs-like potential exerted by an electric field and obtains free energy larger than the amount of work performed on it.
This enables us to verify the generalized Jarzynski equality\cite{Sagawa2010}, or a new fundamental principle of ``information-heat engine'' which converts information to energy by feedback control.
\end{abstract}

\baselineskip 14pt

\maketitle

% Introduction

\begin{figure}
\includegraphics[scale=0.7]{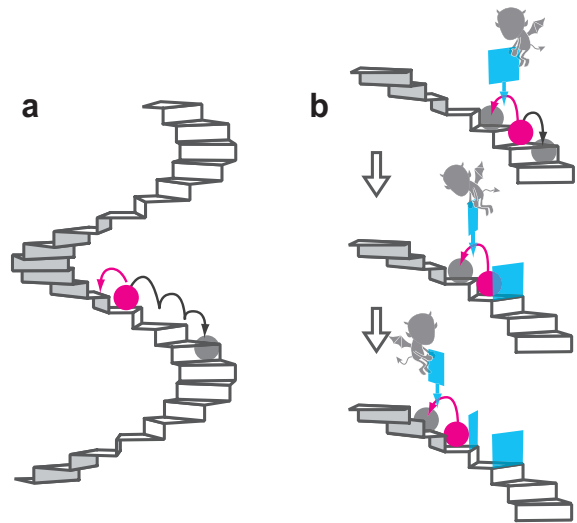}
\caption{\label{fig:Spiral}
Schematic illustration of the experiment.
{\bf a}, A microscopic particle on a spiral-stairs-like potential with a step height comparable to $\kB T$.
The particle stochastically jumps between steps due to thermal fluctuations. 
Since the downward jumps along the gradient are more frequent than the upward ones, the particle falls down the stairs, on average.
{\bf b}, Feedback control.
When an upward jump is observed, a block is placed behind the particle to prevent downward jumps.
By repeating this cycle, the particle is expected to climb up the stairs without direct energy injection.
}
\end{figure}

To illustrate the basic idea of our feedback protocol, let us consider a microscopic particle on a spiral-stairs-like potential (Fig. \ref{fig:Spiral}).
We set the height of each step comparable to the thermal energy $\kB T$, where $\kB$ is the Boltzmann constant and $T$ is the temperature.
Subjected to thermal fluctuations, the particle jumps between steps stochastically.
Although the particle sometimes jumps to an upper step, downward jumps along the gradient are more frequent than upward jumps.
In this manner, on average, the particle falls down the stairs unless it is externally pushed up (Fig. \ref{fig:Spiral}a).
Now, let us consider the following feedback control.
We measure the particle's position at regular intervals, and if an upward jump is observed, we place a block behind the particle to prevent subsequent downward jumps (Fig. \ref{fig:Spiral}b).
If this procedure is repeated, the particle is expected to climb up the stairs.
Note that, in the ideal case, energy to place the block can be negligible; this implies that the particle can obtain free energy without any direct energy injection.
In such a case, what drives the particle to climb up the stairs?
This apparent contradiction to the second law of thermodynamics, epitomized by Maxwell's demon, inspired many physicists to generalize the principles of thermodynamics\cite{Szilard1929, MaxwellDemon2, Maruyama2009}.
It is now understood that the particle is driven by the ``information'' gained by the measurement of the particle's location\cite{MaxwellDemon2, Serreli2007}.\\

In microscopic systems, thermodynamic quantities such as work, heat, and internal energy do not remain constant but fluctuate\cite{Bustamante2005, Sekimoto2010}.
In fact, stochastic violations of the second law have been observed\cite{Wang2002, Carberry2004}; nonetheless, the second law still holds, on average, if the initial state is in the thermal equilibrium : $\langle\Delta F- W\rangle\le 0$, where $\Delta F$ is the free energy difference between states, $W$ the work performed on the system, and $\langle\cdot\rangle$ the ensemble average.
However, the feedback control enables us to selectively manipulate only those fluctuations that violate the second law such as upward jumps by using the information about the system\cite{Thorn2008, Lopez2008, Jourdan2007}.
Our Gedankenexperiment shows that by employing feedback control, the information can be used as a resource for free energy.
In fact, Szilard has developed a model that converts one bit of information about the system to $\kB T \ln 2$ of free energy or work \cite{Szilard1929}. 
In other words, the second law is generalized\cite{Sagawa2008} as follows:
\begin{equation}\label{eq:generalized second law}
\langle\Delta F-W\rangle\le k_\mathrm{B}TI.
\end{equation}
Here $I$ is the mutual information content obtained by measurements\cite{Maruyama2009, Cover1991} (see Methods).
So far, the idea of a simple thermal rectification by feedback control has found applications such as the reduction of thermal noise\cite{Jourdan2007} and the rectification of an atomic current at low temperature\cite{Thorn2008}.
On the other hand, the Szilard-type Maxwell's demon enables us to evaluate both the input (utilized information content) and the output (obatined energy) of the feedback control and relate them operationally.
Therefore, it has provided an ideal test-ground of information-energy conversion and played the crucial role in the foundation of thermodynamics.
However, its experimental realization has been elusive.
In this experiment, we developed a new method to evaluate the information contents and thermodynamic quantities of feedback systems and demonstrate the Szilard-type information-energy conversion for the first time using the colloidal particle on a spiral-stair-like potential.\\

\begin{figure}
\includegraphics[scale=0.8]{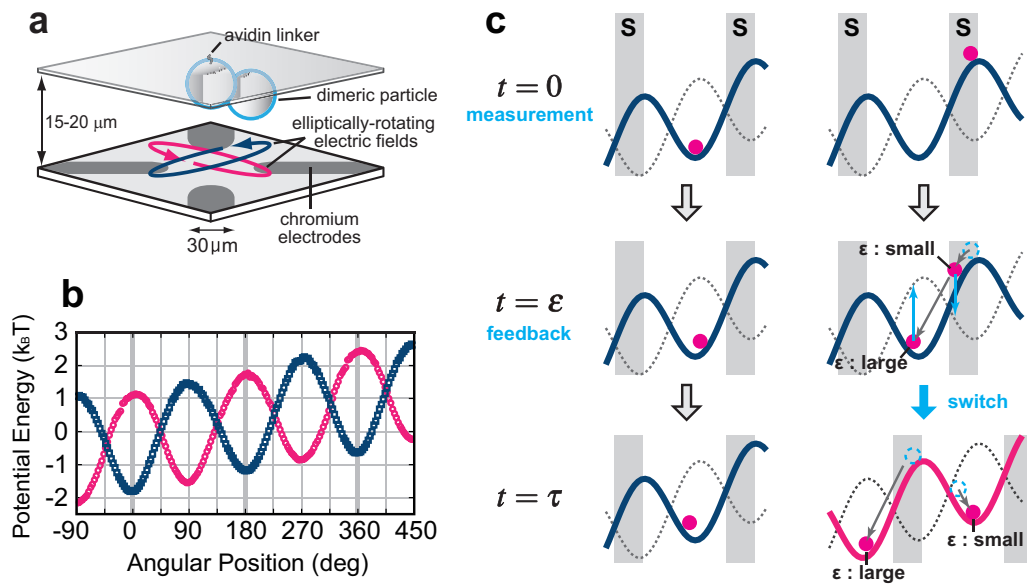}
\caption{\label{fig:System}
Experimental setup\cite{Watanabe-Nakayama2008, Toyabe2010a}. 
{\bf a}, A dimeric particle composed of particles with a diameter of 287 nm (Seradyn) was non-specifically attached to the top glass surface via a streptavidin linker coated on the particle's surface.
The particle was pinned at a single point and exhibited rotational Brownian motion.
To impose a tilted periodic potential on the particle, an elliptically rotating electric field (blue and gray curves) was induced by applying 1 MHz sinusoidal voltages on the quadrant electrodes patterned on the bottom glass surface. 
The particle was observed on a microscope equipped with a high-speed camera at a period of 1.1 ms with an exposure time of 0.3 ms.
Not to scale.
See SI for details.
{\bf b}, Typical potentials with opposite phases to be switched in the feedback control.
Potentials were measured from transition probabilities (see SI).
The particle experienced a tilted periodic potential with a period of 180 degrees.
The direction of the long axis of the elliptically rotating electric field corresponds to the local minima of the potential.
By changing the direction of its axis, we inverted the phase of the potential.
The height and slope were 3.05$\pm$ 0.03 $\mathrm{\kB T}$ and 1.13 $\pm$ 0.06 $\mathrm{\kB T/360^\circ}$ (mean$\pm$S.E., 7 particles), respectively.
{\bf c}, Feedback control. 
At $t=0$, the particle's angular position is measured. 
If the particle is observed in the angular region indicated by ``S,'' we switch the potential at $t=\varepsilon$ by inverting the phase of the potential (right). 
Otherwise, we do nothing (left).
At $t=\tau$, the next cycle starts.
The location of the region S is altered by the switching.
Potential's wells correspond to the steps of the spiral stairs in Fig. \ref{fig:Spiral}.
The switching of potentials corresponds to the placement of the block.  
}
\end{figure}

A dimeric particle comprising polystyrene beads (diameter = 287 nm) was attached to the top glass surface of a chamber filled with a buffer solution(Fig. \ref{fig:System}a).
The particle was pinned at a single point by a linker molecule; it exhibited rotational Brownian motion.
By using quadrant electrodes imprinted on the bottom glass plate, we imposed 1-MHz electric fields to simultaneously create periodic potentials and constant torque on the particle along the angle of rotation. 
By using this new method, a tilted periodic potential with an ideal sinusoidal shape for the particle can be realized, which is a realization of the spiral-stairs-like potential mentioned above (Fig. \ref{fig:System}b, see also SI).
A feedback control was performed under a microscope by constructing a real-time feedback system including video capture, image analysis, potential modulation, and data storage.
We repeated the following feedback cycle with a period of $\tau=44$ ms and a minimum feedback delay of 1.1 ms, as illustrated in Fig. \ref{fig:System}c.
At $t=0$, the particle's angular position is measured.
If the particle is observed at the angular region indicated as ``S,'' the potential is changed to the one with an opposite phase at $t=\varepsilon$; otherwise, no action is taken.
At $t=\tau$, the next cycle begins with the measurement of the angular position.
The region S was chosen for its energy advantage; in the region S, the potential energy before switching is always higher than that after switching.
In the case of small $\varepsilon$, the particle is expected to be at rest around the region S just before the switching at $t=\varepsilon$ and then jump to the rightward well of the switched potential after the switching.
On the other hand, for large $\varepsilon$, the particle falls down in the well away from the region S before the switching.
In this case, with a large probability, the particle jumps down to the leftward well of the switched potential after the switching. 
In this manner, the feedback delay $\varepsilon$ regulates the efficiency of the feedback control.
Note that, since $\tau=44$ ms is sufficiently larger than the relaxation time in each well ($\sim$10 ms) and smaller than the typical time to jump to neighbor wells ($\sim$1 s), each feedback cycle is supposed to be a transition between equilibrium states.\\

\begin{figure}
% 100112/A010.avi
\includegraphics[scale=1]{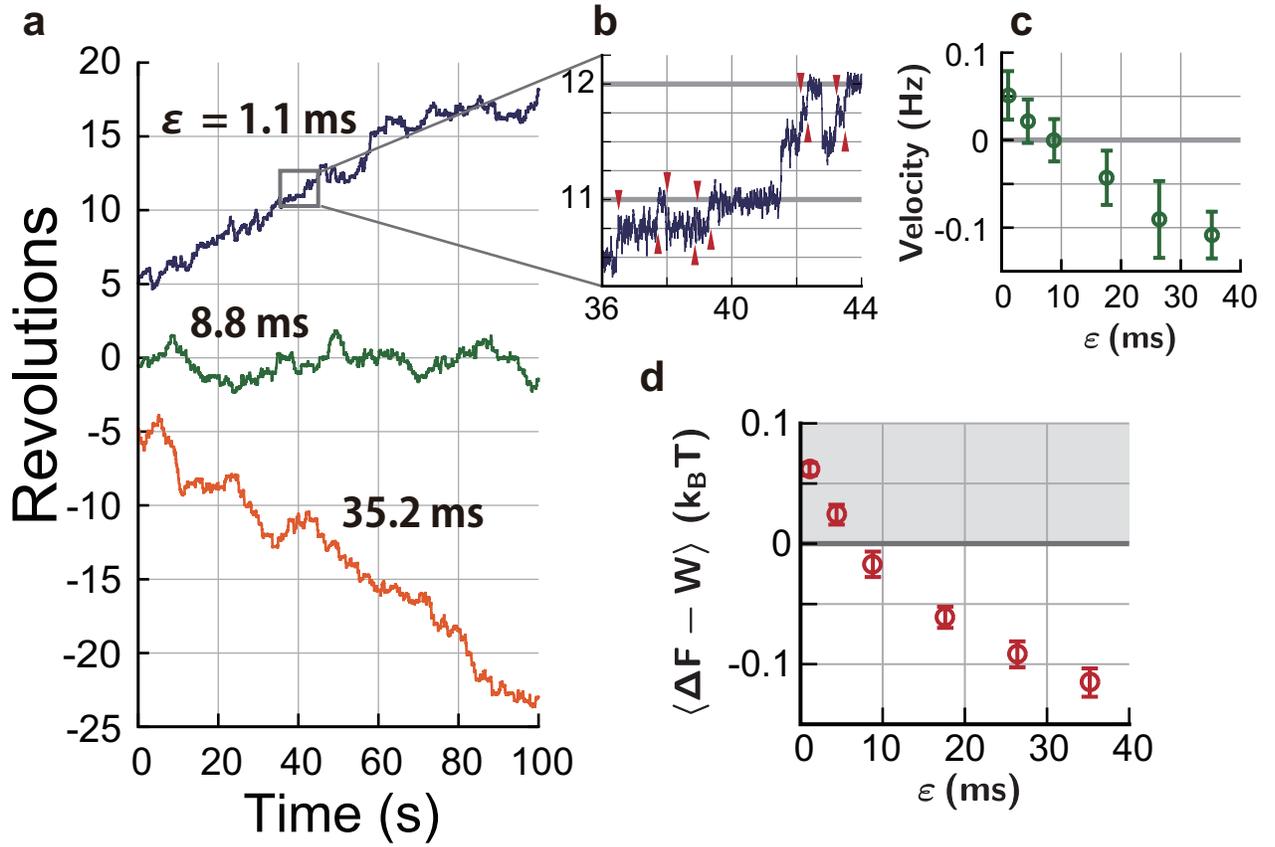}
\caption{\label{fig:Trajectory}
Trajectories, mean velocities, and excess free energy under feedback control.
{\bf a}, Typical trajectories for different values of the feedback delay $\varepsilon$.
{\bf b}, Magnified plot of the region indicated by a rectangle in (a).
The particle rotates with steps with a step size of 90 degrees reflecting the profile of the potential.
Red triangles indicate the timings of the switchings.
{\bf c}, Variation of the rotation rate with feedback delay $\varepsilon$. 
The rotation rate is defined as positive when the particle climbs up the potential. 
Data of seven particles are averaged.
Error bars indicate standard deviations.
{\bf d}, $\Delta F$ : the free energy difference between the initial and final states of the cycle; $W$ : the amount of work performed on the particle by the switching calculated as the potential-energy change associated with the switching (see Methods).
In cycles without switching, $W=0$.
$\langle\cdot\rangle$ denotes the mean per cycle.
In the shaded region, we obtain the excess free energy beyond the conventional limitation of the second law of thermodynamics.
Error bars indicate standard deviations. 
}
\end{figure}

In Fig. \ref{fig:Trajectory}a, typical trajectories with the feedback control are shown.
The trajectories are stepwise with a step size of 90 degrees, which reflects the potential profile (Fig. \ref{fig:Trajectory}b).
We find that for small $\varepsilon$, the particle rotates uni-directionally while climbing up the potential, whereas for large $\varepsilon$, the particle goes down along the gradient.
The rotation rate decreases monotonically with $\varepsilon$, as expected (Fig. \ref{fig:Trajectory}c).\\

We then focused on the energetics during a cycle.
In Fig. \ref{fig:Trajectory}d, we show the difference between the obtained free energy $\Delta F$ and the work performed on the particle by the switching, $W$, which is averaged over a cycle (see Methods). 
We find that $\langle \Delta F-W\rangle > 0$ for small $\varepsilon$; this implies that the particle gains a net free energy larger than the work performed by absorbing heat beyond the conventional limitation of the second law of thermodynamics. 
For small $\varepsilon$, the switching mostly occurs when the particle is in the region S.
In such cases, the particle absorbs heat from an isothermal environment to reach the region S before the measurements at $t=0$, then performs work to the electric field at the switching, and finally jumps to the rightward well after the switching (Fig. S7).
Although such an event is not prohibited even if we randomly switch potentials without feedback control, it is typically an accidental and rare event in accordance with the second law of thermodynamics or the fluctuation theorem\cite{Evans1993, Gallavotti1995, Crooks1999}.
However, the feedback control can increase the likelihood of occurrence of such an event.
This is the crux of the control by Maxwell's demon.
The resource of the excess free energy is the information obtained by the measurement.
If the estimation error of the particle's angular position is negligible, the amount of information is characterized by the Shannon information content $I$ \cite{Shannon1948}.
In this study,  $I=-p\ln p-(1-p)\ln(1-p)$, where $p$ is the probability that the particle is observed in the region S (see Methods).
As noted in (\ref{eq:generalized second law}), $I$ can be converted to free energy of up to $\kB TI$ \cite{Szilard1929}.
In our system, for the shortest feedback delay ($\varepsilon=1.1\,\mathrm{ms}$), $p$, $I$, and $\langle\Delta F-W\rangle$ were 0.059, 0.22, and 0.062\,$\kB T$, respectively (Fig. S9).
This gives the efficiency of the information-energy conversion as, $\langle\Delta F -W\rangle/k_\mathrm{B} TI = 28\%$.
The 100\% efficiency can be achieved by quasi-static information heat engines such as the Szilard engine \cite{Szilard1929}.\\

\begin{figure}
\includegraphics[scale=1]{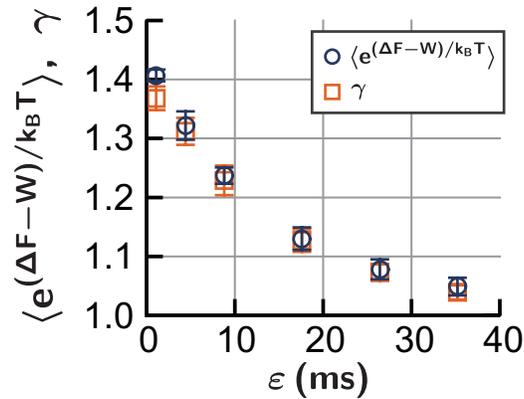}
\caption{\label{fig:Sagawa-Ueda}
Verification of the generalized Jarzynski equality. 
Circles : $\langle e^{(\Delta F-W)/k_\mathrm{B}T}\rangle$. 
Rectangles : feedback efficacy $\gamma$.
Data of seven particles are averaged.
Error bars indicate standard deviations. 
}
\end{figure}

Although the second law concerns only the average, or the first-order cumulant, of the stochastic quantity $\Delta F-W$, Jarzynski pointed out that the second law naturally emerges as the first-order cumulant expansion of the following equality that involves $\Delta F -W$ to all orders \cite{Jarzynski1997, Liphardt2002}: $\langle e^{(\Delta F-W)/\kB T}\rangle = 1$. 
Recently, the Jarzynski equality, which assumes a prescribed control scheme, was generalized to systems with a feedback control as follows\cite{Sagawa2010}:
\begin{equation}\label{eq:Sagawa-Ueda}
\langle e^{(\Delta F-W)/\kB T}\rangle=\gamma,
\end{equation}
where $\gamma$ is an experimentally measurable quantity and defined as the sum of the probabilities that the time-reversed trajectories are observed under time-reversed protocols for all possible protocols (see Methods).
From its definition, $0\le\gamma\le 2$ in our system.
While $I$ concerns the information obtained by the measurements, $\gamma$ quantifies how efficiently we utilize the obtained information for the feedback control.
If we control the system perfectly and deterministically, a time-reversed trajectory is always realized under the time-reversed protocol; $\gamma$ then takes the maximum value.
We repeated time-reversed cycles with/without switchings to obtain $\gamma$ with a period of 220 ms, which is sufficiently long to ensure that the initial state of the cycle is relaxed to equilibrium.
Figure \ref{fig:Sagawa-Ueda} shows that the conventional Jarzynski equality is violated in the presence of the feedback control.
For large $\varepsilon$ where $\langle\Delta F-W\rangle\le 0$, the second law holds on average, but the Jarzynski equality is violated.
On the other hand, the generalized Jarzynski equality (\ref{eq:Sagawa-Ueda}) holds over a broad range of $\varepsilon$, showing that (\ref{eq:Sagawa-Ueda}) expresses the effect of feedback control to all orders.
$\gamma$ appears to converge to 1 in the limit of infinite $\varepsilon$; here, the angular position at the switching becomes independent of that at the measurement, and the conventional Jarzynski equality recovers.
The validity of (\ref{eq:Sagawa-Ueda}) verifies a new fundamental principle of an ``information-heat engine'', which converts information to free energy, in the all-order terms.	\\

We have demonstrated the information-heat engine, which converts information to free energy, as the first realization of a Szilard-type Maxwell's demon.
Note that, since the energy converted from information is compensated for by the demon's energy cost to manipulate information\cite{Landauer1961, Bennet1982, Sagawa2009}, the second law of thermodynamics is not violated when the total system including both the particle and demon is considered.
In our system, the demon comprises of macroscopic devices such as computers, the microscopic device gains energy at the expense of the energy consumption of a macroscopic device.
In other words, by using information as the energy-transferring ``medium'', this information-energy conversion can be utilized to transport energy to nanomachines \cite{Leigh2003, Delden2005} even if it is not possible to drive them directly (Fig. S1).
It is the future challenge to realize a nanoscale information-processing device such as an artificial molecular motor\cite{Hanggi2009, Bier2007}, in which both the demon and controlled system are microscopic.

\section{methods}
%Put methods in here.  If you are going to subsection it, use
%\verb|\subsection| commands.  Methods section should be less than
%800 words and if it is less than 200 words, it can be incorporated
%into the main text.

\subsection{Free energy and work.}

Each potential well separated by peaks was defined as a state (Fig. S5).
The free energy of the state $k$ was caluclated as $F_\mathrm{k}=-\kB T\ln\left[\int\mathrm{d}x\,e^{-U(x)/\kB T}\right]$, where the integration was performed in the angular region corresponding to the state $k$.
Since the shapes of all the wells are almost the same, the free energy difference between states is nearly equal to the difference of the potential energy of their local minima. 
The work performed on the particle $W$ was calculated as the potential-energy change associated with the switching; the potential energy after the switching minus that before the switching. 
In cycles without switching, $W=0$.

\subsection{Information content.}

For an event $k$ with a probability of occurrence $p(k)$, the Shannon information content associated with this event is defined as $-\ln p(k)$.
This definition leads to well-defined properties that the information content should satisfy\cite{Shannon1948}.
The average Shannon information content becomes $I\equiv -\sum_k p(k)\ln p(k)$.
Measurements usually accompany errors, which reduce the amount of information that can be utilized.
Although $I$ denotes the amount of the information embedded in the system, the mutual information content, $I'$, denotes the amount of information that is obtained by the measurement \cite{Cover1991, Sagawa2010}: $I'\equiv \sum_{k,m} p(m|k)p(k)\ln\frac{p(m|k)}{p(k)}$, where $p(m|k)$ is the conditional probability that the outcome of the measurement is the $m$-th event when the $k$-th event occurs actually.
If the measurement is free from error, $p(m|k)=\delta_\mathrm{k, m}$ ($\delta_{k, m}=1$ if $k=m$, and otherwise 0).
In such case, $I'=I$.
In the present experiment, we distinguished two events; the particle is observed in the region S or not with negligible measurement errors.
Then, the (average) Shannon information content per cycle becomes the so-called binary entropy function: $I= -p\ln p-(1-p)\ln(1-p)$, where $p$ is the probability that the particle is observed in the region S.

\subsection{Feedback efficacy.}

The feedback efficacy $\gamma$ is defined as the sum of the probabilities that the time-reversed trajectories are observed under time-reversed protocols for all possible protocols (Fig. S7).
In the forward feedback cycle, we measure the particle's angular position at $t=0$ and (i) switched or (ii) did not switch the potential at $t=\varepsilon$ depending on the angular position.
Corresponding time-reversed trajectories are that the particle is observed in the region (i) S at $t=\tau$ after the switching at $t=\tau-\varepsilon$ or (ii) outside S without switching.
Let the occurrence probabilities of time-reversed trajectories under timer-reversed protocolos be $p_\mathrm{sw}$ and $p_\mathrm{ns}$, respectively.
Then, $\gamma$ is $\gamma=p_\mathrm{sw}+p_\mathrm{ns}$.
From its definition, if there are $m$ states to be distinguished ($m=2$ in our experiment: whether the particle is in the region S or not), $0\le\gamma\le m$.
We repeated time reversed cycles with/without switchings to obtain $\gamma$ with a period of 220 ms, which is sufficiently long to ensure that the initial state of the cycle is relaxed to equilibrium.

\bibliography{manuscript}

\begin{description}
 \item[Acknowledgements] We thank  T. Okamoto, M. Miyazaki, and N. Akiyama for critical discussions and A. Ozeki for illustrations. This work was supported by Japan Science and Technology Agency (JST) and Grant-in-Aid for Scientific Research, 18074001, 17049015, 19037022, 18031033 (to EM), and 21740291, 21023007 (to ST). TS acknowledges JSPS Research Fellowships for Young Scientists (208038). \item[Author contributions] S.T. designed and performed experiments, analysed data and wrote the paper. T.S. and M.S. designed experiments and wrote the paper. T.S. and M.U. supported theoretical aspects. E.M. and M.S. supervised the experiments. All authors discussed the results and implications and commented on the manuscript at all stages.
 \item[Author Information] The authors declare that they have no competing financial interests. Correspondence and requests for materials should be addressed to M.S. (sano@phys.s.u-tokyo.ac.jp) or E.M. (emuneyuk@phys.chuo-u.ac.jp).
\end{description}

\newpage

% Figures

\end{document}

\title{Supplementary Information for\\ "Information heat engine: converting information to energy by feedback control"}

\author{Shoichi Toyabe}
%\email{toyabe@phys.chuo-u.ac.jp}
\affiliation{\chuo}
\author{Takahiro Sagawa}
\affiliation{\tokyo}
\author{Masahito Ueda}
\affiliation{\tokyo}
\affiliation{\ERATO}
\author{Eiro Muneyuki}
\affiliation{\chuo}
\author{Masaki Sano}
\affiliation{\tokyo}

\maketitle

\hrule

%\tableofcontents

\baselineskip14pt

\section{Schematic of experiment and the consistency with the second law of thermodynamics.}

We demonstrated that free energy is obtained by a feedback control using the information about the system; information is converted to free energy, as the first realization of Szilard-type Maxwell's demon\cite{Szilard1929}.

\begin{figure}[htb]
\centering{\includegraphics[scale=0.6]{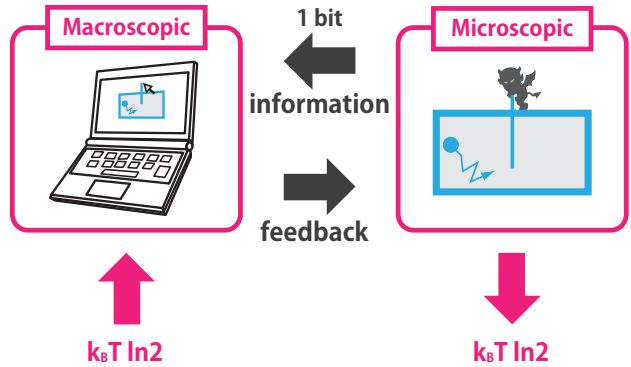}}
\caption{Schematic of the information-energy conversion with a macroscopic demon and a microscopic system (Szilard engine\cite{Szilard1929} as an example).
The Szilard-engine can achieve the 100\% conversion rate from information to energy; 1 bit ($\ln 2$ nat in the natural logarithm) of information is converted to free energy or work of $k_\mathrm{B}T\ln 2$ in the microscopic system at the expense of an energy consumption of $k_\mathrm{B}T\ln 2$ in the macroscopic demon.
}
\label{fig:Second Law}
\end{figure}

Since the obtained free energy or work is compensated for by the demon's energy cost to manipulate information, it does not violate the second law of thermodynamics when the total system including both the particle and demon is considered.
The demon consists of macroscopic devices in practice such as computers in our system; the microscopic device gains energy at the expense of the energy consumption of a macroscopic device (Fig. S\ref{fig:Second Law}).
In other words, by using information as the energy-transferring ``medium'', this information-energy conversion can be utilized to transport energy to nanomachines even if it is not possible to drive them directly.

\section{Preparation of chamber.}

Brownian particles in a buffer solution were prepared for the experiment by the following procedure. 
The cover slip (Matsunami, Japan) that served as the ceiling of the observation chamber was washed three times with distilled water.
The bottom glass coated with chromium electrodes (Matsunami, Japan) was soaked in 10 N KOH over night and then washed with distilled water.
The top and bottom glasses were separated by a plastic film with a thickness of about 12 $\mu$m (Saran wrap, Asahi Kasei, Japan) and silicone grease.
6 $\mu$l of A buffer (50 mM MOPS-K, 50 mM KCl, 2 mM MgCl$_2$, 2 mg/ml heat-shocked Bovine serum albumin (Sigma-Aldrich, USA), pH6.9) was applied to the chamber to suppress the binding of particles to the glass surface.
Then, 10 $\mu$l of streptavidin-coated polystyrene beads with a diameter of 287 nm (Seradyn, USA) in A buffer was applied to the chamber.
Some particles bound to the top glass surface.
About 50 $\mu$l of B buffer (5 mM MOPS-K, 1 mM MgCl$_2$, 1 mM KP$_\mathrm{i}$, pH6.9) was applied to wash out unbound particles in the chamber.
All the experiments were performed in B buffer at a room temperature.

\section{Microscopy and image analysis.}

The incidentally-dimeric particle pinned at the top glass surface was observed on a phase-contrast microscope (BX-51WI, Olympus, Japan) equipped with a high speed camera (IPX-210L, Imperx, USA) at a period of 1.1 ms with an exposure time of 0.3 ms.
Real-time feedback system was constructed on the PXI (PCI eXtensions for Instrumentation) system implementing a computing module (PXI-8108, National Instruments, USA), a video capturing board (PXI-1428, National Instruments), and a digital-to-analog conversion board for the control of the electric potential across the electrodes patterned on the bottom glass plate (PXI-6221, National Instruments).
The minimum delay in this feedback control is thus 1.1 ms  which is much shorter than the time for the particle to perform rotational diffusion up to 90 degrees.
The angle of the dimeric probe particle was estimated using an algorithm based on the principal component analysis.

\section{Elliptically-rotating electric field.}

\begin{figure}[htb]
\centering{\includegraphics[scale=1.0]{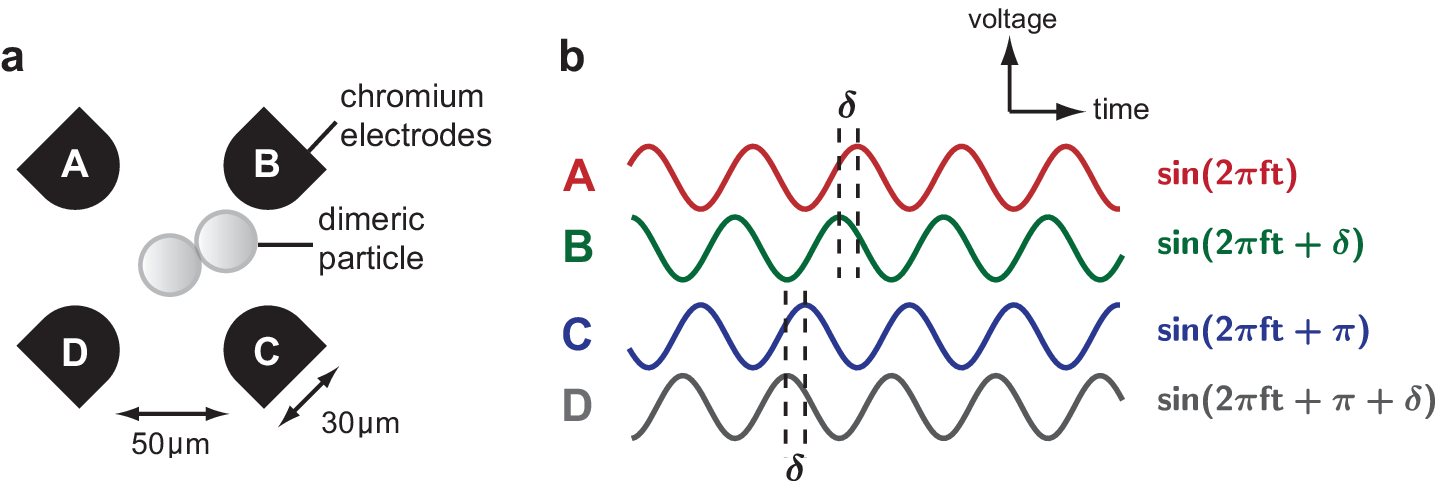}}
\caption{(a) Quadrant electrodes. Not to scale. (b) Sinusoidal voltages applied on the electrodes.}
\label{fig:System}
\end{figure}

\begin{figure}[htb]%
\centering{\includegraphics[scale=1.0]{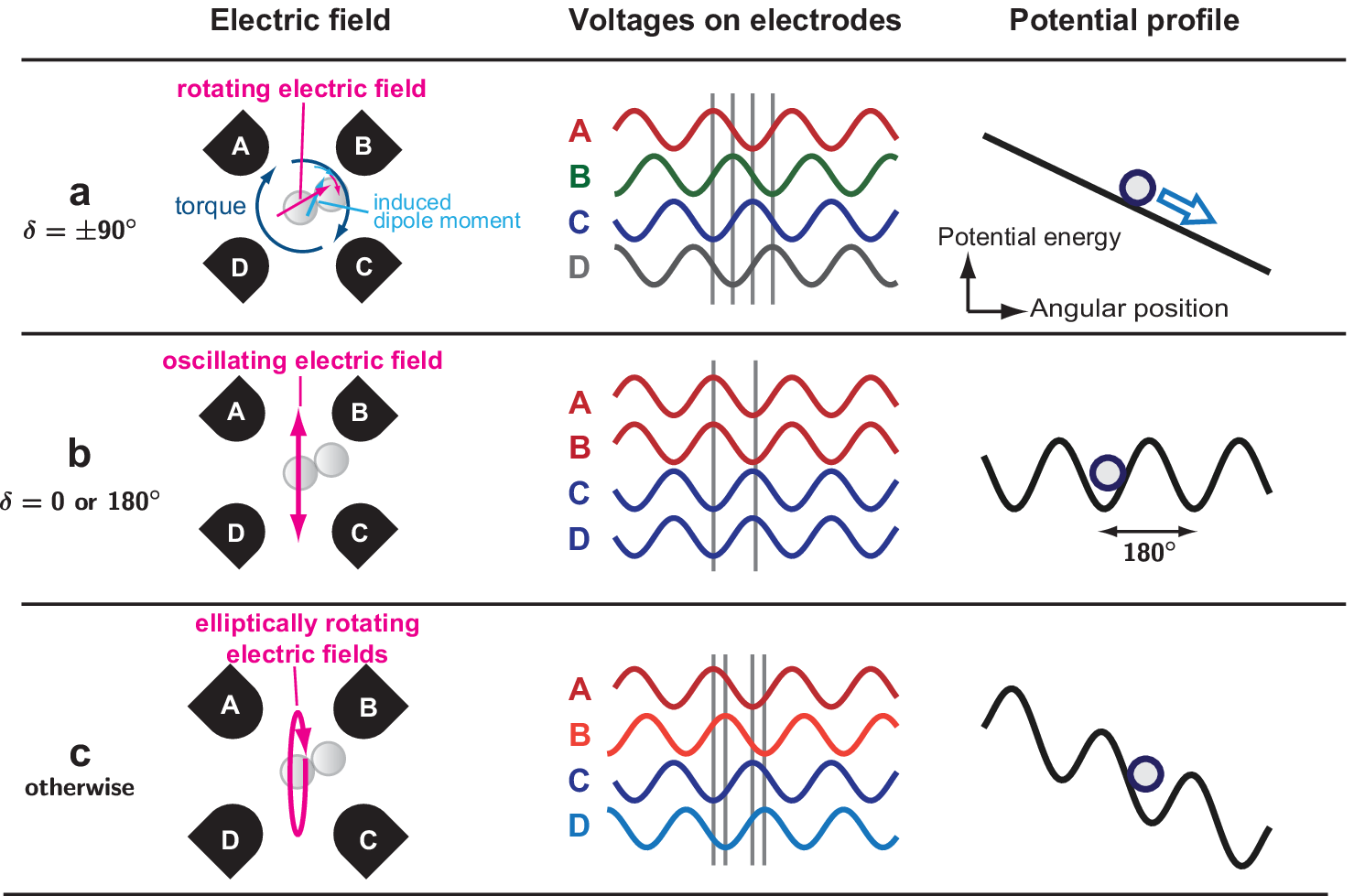}}%
\caption{
Three different types of the potential imposed on the particle according to the phase shift $\delta$.
(a) Rotating electric field.
(b) Oscillating electric field.
(c) Elliptically rotating electric field.
}
\label{fig:EREF}%
\end{figure}

To provide a periodic potential and adverse torque to the Brownian particle, we utilized the induced polarization effect of the dielectric material subjected to an AC external electric field\cite{Watanabe-Nakayama2008, Washizu1993, Toyabe2010a}.
Four chromium electrodes on the bottom glass surface were connected to four power amplifiers (HSA4101, NF, Japan) that amplify AC (Alternative Current) voltages generated by function generators (WF1974, NF, Japan).
We applied 1-MHz AC voltages with some phase shifts on the electrodes (Fig. S\ref{fig:System}).
Let the phases on the electrodes to be 0, $\delta$, 180, and $180+\delta$ degrees.
Three different types of potentials are imposed on the particle according to the value of $\delta$ (Fig. S\ref{fig:EREF}).
\begin{itemize}
\item If $\delta=\pm 90$ degrees, a rotating electric field is applied at the center of the electrodes (Fig. S\ref{fig:EREF}a), where a dipole moment rotating at the same rate with some time delay appears on dielectric particles.
The length of the time delay is determined by the complex dielectric constants of the particle and the solution.
Due to the time delay, the particle experiences a torque from the electric field \cite{Watanabe-Nakayama2008, Washizu1993, Toyabe2010a}.
The torque is constant; it does not depend on the particle's motions but depends only on the time delay.
Its magnitude is proportional to the particles's volume and the square of the amplitude of the applied AC voltages.
Also, it nonlinearly depends on the frequency of the applied AC voltages.
According to the sign of $\delta$, the direction of the torque is changed.
\item If $\delta=0$ or $180$ degrees, the electric field does not rotate but oscillates along an axis (Fig. S\ref{fig:EREF}b).
In this case, the particle feels a periodic potential with a period of 180 degrees along the direction of rotation.
The oscillating direction of the electric field corresponds to the angle of the potential's local minima.
\item Otherwise, an elliptically rotating electric field is applied (Fig. S\ref{fig:EREF}c); it corresponds to the superposition of the above two cases ($\delta=0$ and 90 degrees).
In such case, the particle feels a tilted periodic potential.
The gradient of the potential depends on the value of $\delta$. 
According to the sign of $\delta$, the direction of the torque is changed.
\end{itemize}

We set $\delta$ to be about 5 degrees ($\delta$ was tuned for each particle) to apply the elliptically rotating electric field.
For feedback control, the potential was switched by modulating the phases of applied voltages from 0, $\delta$, 180, and 180+$\delta$ degrees to 0, 180-$\delta$, 180, and -$\delta$ degrees, respectively for four electrodes.

\section{Estimation of potential energy.}

\begin{figure}[htb]%
\centering{\includegraphics{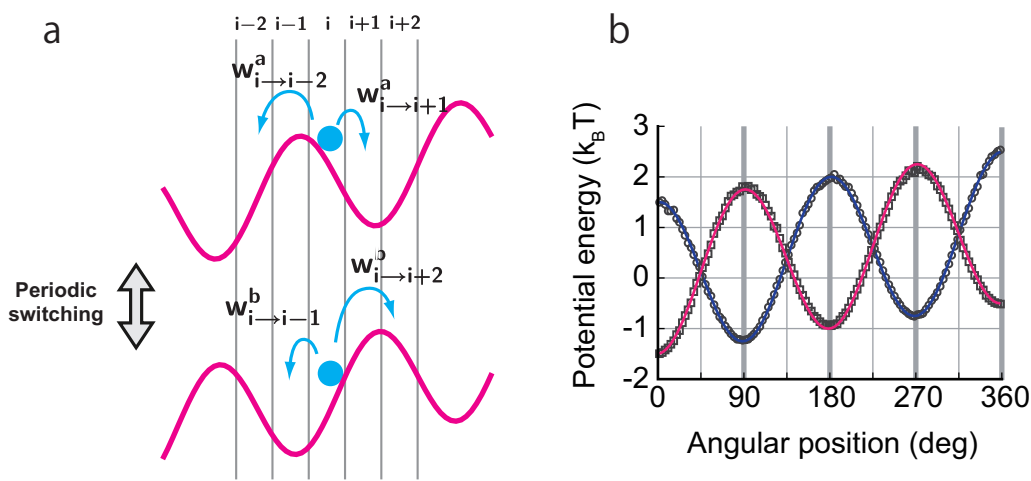}}%
\caption{
(a) Potential estimation.
(b) Estimation of the potential energy for a simulated trajectory.
Solid lines indicate the theoretical curves.
Markers correspond to estimated values (bin width = 3 degrees).
The slope is 1 $k_\mathrm{B}T/360^\circ$, the activation energy (peak-to-peak amplitude) 3 $k_\mathrm{B}T$, the rotational friction coefficient 0.3 pN$\cdot$nm$\cdot$s/rad$^2$, the framerate 1000 Hz, and temperature 300 K.
Number of frames is 1,000,000.
The potentials were switched every 200 frames.
}
\label{fig:PotentialEstimation}%
\end{figure}

The potential profiles were estimated by a method similar to Ref. \cite{Errington2004} (Fig. S\ref{fig:PotentialEstimation}a). 
We switched two potentials with opposite phases periodically with a period of 220 ms to sample data around not only the local minima but also the peaks.
We divided the whole video frames to two sets according to the potential at each frame.
For each set of frames, we performed the following procedures to estimate the potential profile.
360 degrees are divided to equally-spaced bins with a bin width of 3 degrees.
We counted the number of video frames $m_\mathrm{i}$ where the particle is in the $i$\,th bin.
We also counted the number of transitions $n_\mathrm{i\to j}$, or the number of two-successive-frame where the particle in the $i$\,th bin moves to the $j$\,th bin at the next frame.
The transition probabilities $w_\mathrm{i\to j}\equiv n_\mathrm{i\to j}/m_\mathrm{i}$ satisfy the detailed balance condition:
\begin{equation}\label{eq:Detailed Balance Condition}
\frac{w_\mathrm{i\to j}}{w_\mathrm{j\to i}}=e^{-\Delta U_\mathrm{i\to j}/\kB T},
\end{equation}
where $\Delta U_{i\to j}\equiv U_\mathrm{j}-U_\mathrm{i}$ is the difference of the potential energies between $i$\,th and $j$\,th bins.
We searched the set $\{U^*_i\}$ which minimizes a score function as follows:
\[
\varepsilon^2(\{U_i\})\equiv \sum_{i<j}\sqrt{n_\mathrm{i\to j}n_\mathrm{j\to i}}\left[\Delta U_\mathrm{i\to j}-\Delta U'_\mathrm{i\to j}\right]^2,
\]
where $\Delta U'_\mathrm{i\to j}\equiv\kB T \left[\ln{w_\mathrm{j\to i}}-\ln{w_\mathrm{i\to j}}\right]$, using the Powell's method \cite{NumericalRecipes3}.

The validity of this method was verified for numerically-simulated trajectories based on a Langevin dynamics.
Figure S\ref{fig:PotentialEstimation}b shows that this algorithm estimates the potential energy for the trajectory generated by a Langevin simulation to a good extent.

A typical number of video frames used for the potential estimation was about 1,000,000.

\section{State and free energy.}

\begin{figure}[htb]
\centering{\includegraphics[scale=1.0]{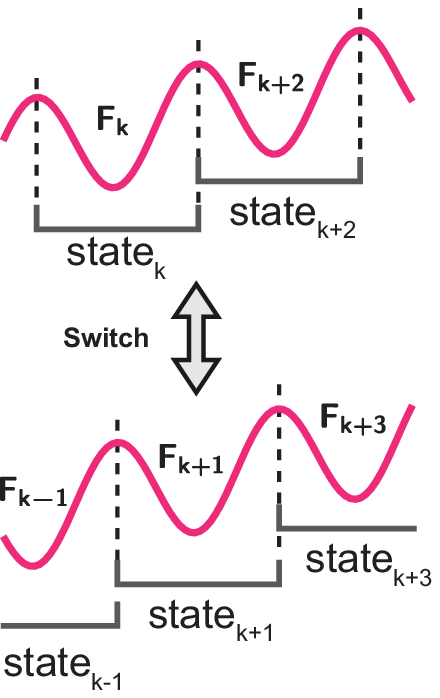}}
\caption{Definitions of the states and free energies. The well separated by peaks is regarded as a state. A free energy is assigned to each state.}
\label{fig:FreeEnergy}
\end{figure}

Figure S\ref{fig:FreeEnergy} is the schematic to illustrate the definition of the state and how to calculate the free energy.

\section{Schematic of Energetics.}

\begin{figure}[htb]%
\centering{\includegraphics[scale=1.0]{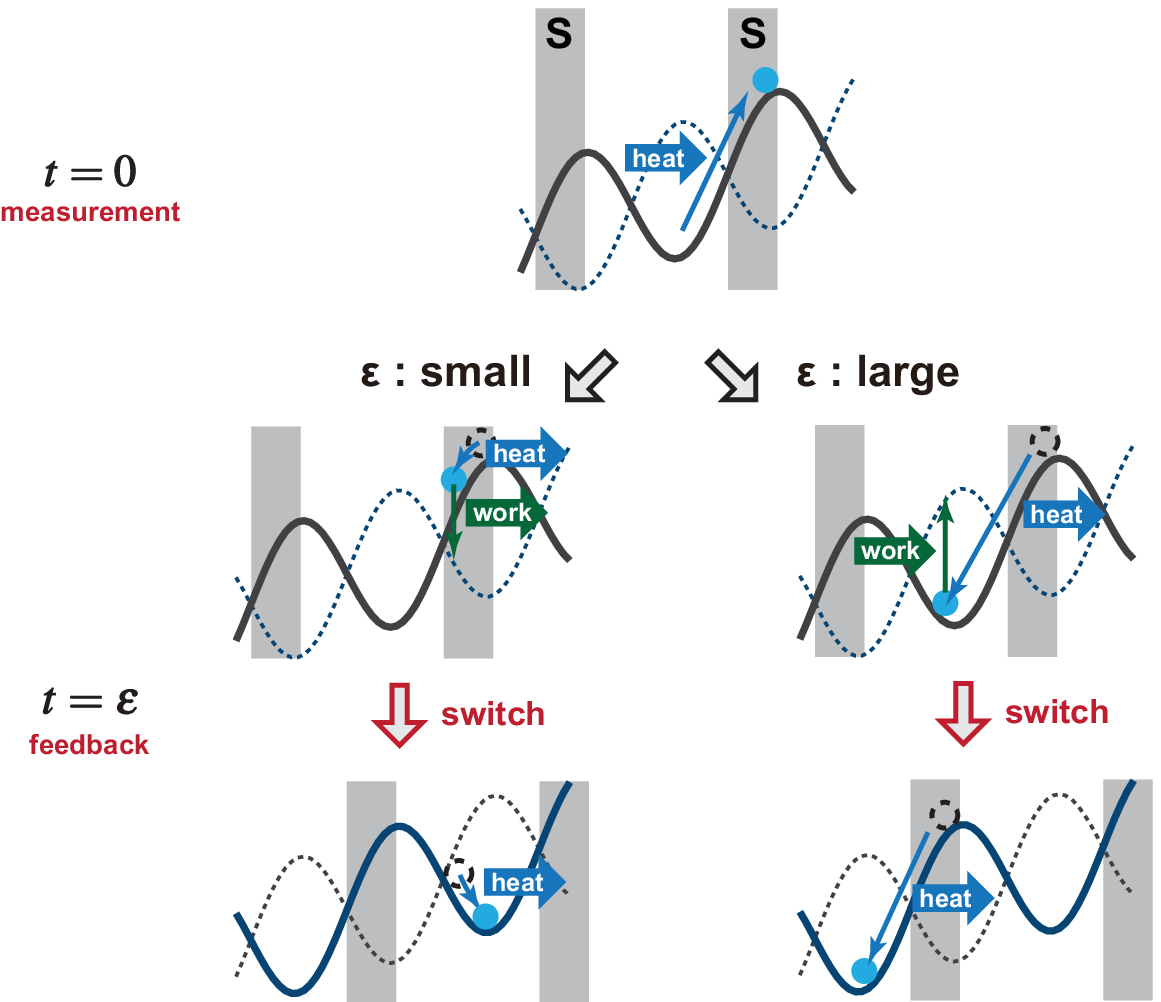}}%
\caption{
Schematic to illustrate the energetics in a cycle with a switching.
For small $\varepsilon$, switchings occur mostly when the particle is in the region S.
In this case, the particle absorbs heat from an isothermal environment to reach the region S before measurements at $t=0$, performs extractable work to the electric field at the switching, and finally jumps to the rightward well after the switching.
On the other hand, for large $\varepsilon$, most of the energy at $t=0$ dissipates to the heat bath before switching.
Then, the switching performs external work on the particle with lifting up the potential, which then dissipates as it falls down in the well.
}
\label{fig:Energetics}%
\end{figure}

Schematic to illustrate the energetics in a cycle with a switching is shown in Fig. S\ref{fig:Energetics}.

\section{Feedback efficacy.}

\begin{figure}[htb]%
\centering{\includegraphics[scale=0.7]{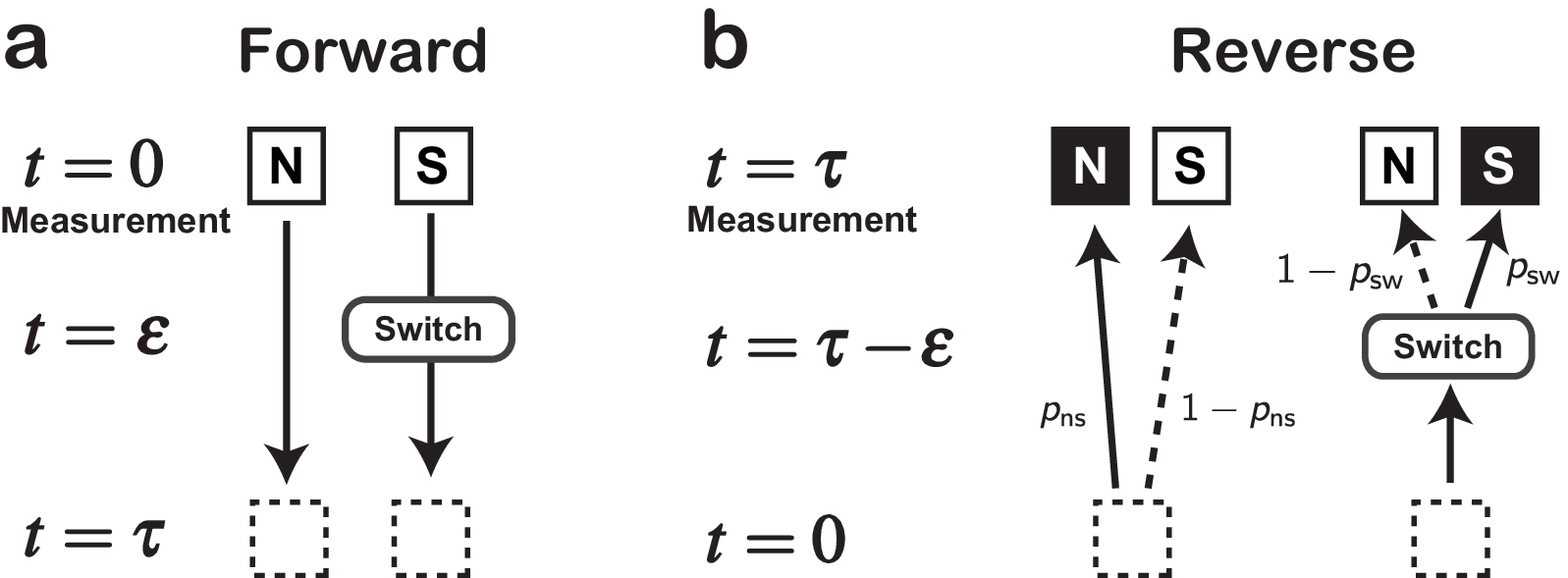}}%
\caption{Forward and reversed cycle. 
(A) Forward feedback cycle. ``S'' and ``N'' indicate that the particle is observed in the region S or not, respectively. 
(B) Reversed cycle. Feedback control is not performed.}
\label{fig:Reverse}%
\end{figure}

Figure S\ref{fig:Reverse} is the schematic to illustrate the feedback efficacy.

\section{Efficiency.}

\begin{figure}[htb]
\centering{\includegraphics[scale=1.0]{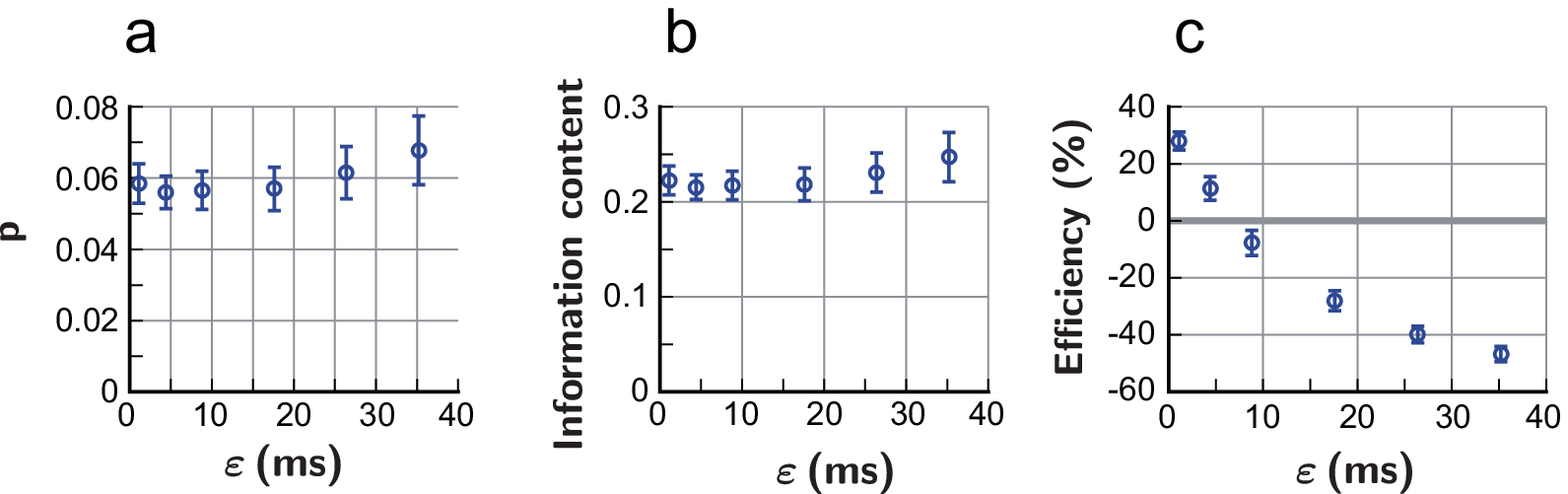}}
\caption{Efficiency of information-energy conversion. 
(a) The probability $p$ that the particle is observed in the region S. 
(b) Shannon information content $I\equiv -p\ln p-(1-p)\ln(1-p)$.
(c) Efficiency of the information-energy conversion calculated as $\langle\Delta F-W\rangle/k_\mathrm{B}TI$.
Error bars indicate standard deviations.
}
\label{fig:Efficiency}
\end{figure}

Figure S\ref{fig:Efficiency} shows (a) the probability that the particle is observed in the region S, (b) the (average) Shannon information content, and (c) the efficiency to convert information to energy.

\bibliographystyle{prsty}
\bibliography{SI}